\documentclass[12pt]{article}
\usepackage{amsmath}
\usepackage{graphicx}
\usepackage{natbib}
\usepackage{url} 

\DeclareMathOperator*{\argmax}{arg\,max}

\usepackage{booktabs}   
\usepackage{pifont}

\newcommand{\blind}{0}

\begin{document}

\def\spacingset#1{\renewcommand{\baselinestretch}%
{#1}\small\normalsize} \spacingset{1}


\if0\blind
{
  \title{\bf Consistent Estimation of  Propensity Score Functions with Oversampled Exposed Subjects}
  \author{Sherri Rose\thanks{
    The author gratefully acknowledges support from the Laura and John Arnold Foundation.}\hspace{.2cm}\\
    Department of Health Care Policy, Harvard Medical School}
  \maketitle
} \fi

\if1\blind
{
  \bigskip
  \bigskip
  \bigskip
  \begin{center}
    {\LARGE\bf Title}
\end{center}
  \medskip
} \fi

\bigskip
\begin{abstract}
Observational cohort studies with oversampled exposed subjects are typically implemented to understand the causal effect of a rare exposure. Because the distribution of exposed subjects in the sample differs from the source population, estimation of a propensity score function (i.e., probability of exposure given baseline covariates) targets a nonparametrically nonidentifiable parameter. Consistent estimation of propensity score functions is an important component of various causal inference estimators, including double robust machine learning and inverse probability weighted estimators. This paper develops the use of the probability of exposure from the source population in a flexible computational implementation  that can be used with any algorithm that allows observation weighting to produce consistent estimators of propensity score functions. Simulation studies and a hypothetical health policy intervention data analysis demonstrate low empirical bias and variance for these  propensity score function estimators with observation weights in finite samples.
\end{abstract}

\noindent%
\vfill

\newpage
\spacingset{1.5} 
\section{Introduction}
\label{intro}
Observational cohort studies with oversampled exposed subjects are performed, in particular, in public health and health policy research to evaluate causal effects.  This design allows  researchers to study multiple outcomes for rare exposures (e.g., Agent Orange in Vietnam, policies implemented in small geographic regions relative to the target population) or exposures that are difficult to measure \citep{Rothman:Greenland98}. A consequence of the conditional sampling is that it creates cohorts where the distribution of exposed subjects differs from the source population.  Consistent estimation of the response surface is not impacted by this oversampling of exposed subjects. However, the propensity score function (i.e., probability of exposure given baseline covariates) targets a nonparametrically nonidentifiable parameter. 

Empirical causal inference involves both the positing of a causal model as well as  estimation techniques that rely on additional statistical assumptions. Establishing causal identifiability, via structural causal models or the Neyman-Rubin causal model, is covered extensively in other literature \citep{Holland86,Neyman23,Pearl09,Rosenbaum02,Rubin74,Rubin06}, and this paper's contribution  centers on the second part of empirical causality. However, this short work does not propose causal estimators, but rather a consistent estimation technique for  propensity score functions in cohort studies with oversampled exposed subjects that relies on a straightforward computational implementation: using observation weights. 

Many estimators in the observational causal inference literature leverage the propensity score function as a nuisance parameter to target effect parameters in an effort to control for confounding.  These include estimating equation and maximum likelihood approaches, such as inverse probability weighted estimators (IPWs) and double robust machine learning estimators \citep{Robins:Hernan:Brumback00,vanderLaan:Rubin06,Cole:Hernan08,vanderLaan:Rose11,chernozhukov2018double}. IPWs depend on consistent estimation of the propensity score function to estimate the target parameter consistently. Double robust machine learning estimators like targeted maximum likelihood estimation require consistent estimation of the response surface \textit{or} the propensity score function for consistent estimation of the target parameter, and will be asymptotically efficient if both are estimated consistently. Thus, a flexible and consistent (possibly machine-learning-based) estimator of the propensity score function for cohort studies with oversampled exposed subjects has multiple potential uses.

This article does not consider cohort studies where unexposed subjects are matched to exposed subjects conditional on particular observed baseline covariates (e.g., age), although this extension is also possible. Nor is the focus studies where the propensity score function is used only in the \textit{design phase} to match unexposed subjects to exposed subjects, rather than as a nuisance parameter  in the estimator after the sample has been constructed.  Matched cohort studies and related techniques, such as synthetic controls, are described in other literature \citep[e.g.,][]{Holland:Rubin88,Rothman:Greenland98,abadie2010synthetic,ryan2015we,athey2017state}.

The observation weights   developed here for  cohorts with oversampled exposed subjects are created using known (or estimated) probability of  exposure from the source population. This concept is similar to the use of known or estimated prevalence probability of disease for case-control studies.  In case-control studies, subjects are sampled conditional on the \textit{outcome} rather than the exposure.  The prevalence probability has been used in case-control studies to develop consistent estimators of (\textit{a}) the response surface in parametric regression using intercept adjustment, (\textit{b}) the response surface with machine learning and observation weights for prediction, (\textit{c}) the response surface and propensity score function with observation-weighted double robust machine learning, and (\textit{d}) predictiveness curves with parametric regression \citep{Anderson72,Prentice:Breslow78,Greenland81,Rose:vanderLaan08tr,Rose:vanderLaan08,huang2009semiparametric}.

This work contributes a  method that yields  consistent estimation of propensity score functions
via observation weights based on the known probability of  exposure from the source population for cohorts with oversampled exposed subjects. A key feature of this approach is that it can be flexibly integrated into computational implementations of  parametric regression and a variety of other existing algorithms for prediction (including stacking).  It is not married to a particular choice of algorithm, other than it allowing for observation weights.  We provide empirical demonstrations of the methodology in simulations, including sensitivity analyses where the probability of exposure is not known but estimated, and a hypothetical health policy intervention data analysis.

\section{Statistical Estimation Problem}

The challenge of the statistical estimation problem for the propensity score function is  to account for the bias imposed by sampling subjects  conditional on  exposure $E$. (We refer to both traditional epidemiologic exposures and policy interventions as `exposures' for simplicity.) Without loss of generality, we consider binary $E \in \{0,1\}$ and a vector of baseline covariates $\mathbf{X}$.  We sample from a conditional distribution of $\mathbf{X}$ given that $E=1$ and  the conditional distribution of $\mathbf{X}$ given  $E=0$.  Thus, the observed data $O$ in a cohort study with oversampled exposed subjects is given by \[O=(\textbf{X}_{E=1},(\textbf{X}^c_{E=0}: c=1,\ldots, C))\sim P\]
\[\textbf{X}_{E=1} \sim (\textbf{X}\mid E=1)\]
\[\textbf{X}^c_{E=0} \sim (\textbf{X}\mid E=0),\]
where $C$ is the number of unexposed `control' subjects per exposed subject and $P$ is a probability distribution  contained within nonparametric model $\cal{M}$.

If  observations were instead sampled from an underlying probability distribution $P$ at random, without conditioning on $E$, our observed data structure would be defined as  \[O^F=(\mathbf{X}, E)\sim P^F\in \cal{M}^F,\] where  the $F$ superscript represents drawing from the `full' data distribution $P^F$ of interest.  While our observed data structure $O$ drawn from $P \in \cal{M}$ differs from $O^F\sim P^F \in \cal{M}^F$, we present methodology in Section~\ref{meth} that allows us to estimate parameters defined by $P^F$. Our target parameter is the conditional distribution: \[\psi=P^F(E=1\mid \textbf{X}),\] which we also refer to as the propensity score function.

In practice, $C$ is the average number of unexposed `control' subjects  per exposed subject in the sample, is permitted to vary, and need not be an integer. We sample $n$ exposed subjects and $n\times C$ unexposed subjects, with $n+(n\times C)=N$ as our total sample size. The outcome variable is not principally relevant to the estimation of the propensity score function, which is why it does not receive formal treatment here. 

\section{Methodology and Computational Implementation}\label{meth}

The premise of our methodology for  propensity score functions relies on knowledge of the probability of  exposure in the source population:
$w=P^F(E=1)$. This value might  be obtained, for example, from registry databases, census information, or other regulatory sources.  When exposed subjects are oversampled in a cohort, estimating the propensity score function in the sample targets a parameter that is not identifiable. However, we propose the use of $w$, such that $\psi=P^F(E=1\mid \textbf{X})$ becomes identifiable, and we have the estimator:
\[\hat{\psi}= \argmax_{\psi} \sum^n_{i=1}\bigg[w \log \hat{P}^F(E_i\mid \textbf{X}_{E=1,i})  + (1-w)\frac{1}{C}\sum^C_{c=1}  \log \hat{P}^F(E_i\mid \textbf{X}^c_{E=0,i})    \bigg].
\]
Here, exposed subjects receive observation weights $w$ and unexposed subjects receive observation weights $(1-w)\frac{1}{C}$.  This form of  weighting to account for the conditional sampling of exposed subjects is quite flexible, as many statistical learning methods accommodate observation weights. It is also agnostic to the choice of loss function. We additionally posit that an approximation of $w$ (e.g., estimated in other studies drawn from the source population) can be used to estimate $\psi$.

Stacking (alternatively, ensembling) is gaining in popularity in substantive research as it permits the application of multiple algorithms without the need to select a single algorithm \textit{a priori} and can accommodate high-dimensional data settings where parametric model misspecification is a substantial concern \citep{Stone74,Geisser75,Wolpert92,Breiman96,Leblanc96,vanderLaan:Polley:Hubbard07}.  We describe the use of the probability of exposure $w$ in the super learner ensemble framework due to its optimality properties \citep{vanderLaan:Polley:Hubbard07}, and to demonstrate the flexibility of our weighting approach in multiple algorithms. This general algorithm is detailed below.

\begin{table}[h]
 \small
  \label{sample-table}
  \centering
  \begin{tabular}{rl}
    \toprule
    \multicolumn{2}{l}{\textbf{Algorithm}  Stacked Propensity Score Function with Weighting}                   \\
    \midrule
    \textit{Input:} & $a$) Observed realizations of cohort data $O=(\textbf{X}_{E=1},(\textbf{X}^c_{E=0}: c=1,\ldots, C))$, \\
    & $b$) probability of exposure $w=P^F(E=1)$, $c$) set of learners, and  $d$) loss function.\\
    \cmidrule(lr){2-2}
    \ding{182} &Assign observation weights $w$ to exposed subjects and $(1-w)\frac{1}{C}$ to unexposed  \\
    &  control subjects. \\
        \ding{183} &Fit all learners using observation weights in $k$-fold cross validation; store cross- \\
        &validated predicted values.\\
            \ding{184} & Regress $E$ on the cross-validated predicted values (using observation weights), \\
            & which minimizes (or maximizes) the chosen loss function and returns   vector of  \\
            & algorithm weights for the stacked ensemble.\\
             \ding{185} & Combine set of learners with fixed estimated vector of algorithm weights; fit entire\\
             &  cohort sample.  \\
             \cmidrule(lr){2-2}
     \textit{Output:} &  \textit{I})  Estimated propensity score function $\hat{\psi}=P^F(E=1\mid \textbf{X})$ and \textit{II})  final estimated\\
     &  predicted values.\\
          \cmidrule(lr){2-2}
            \ding{186} & Cross-validate the entire stacking procedure to obtain  cross-validated  predicted\\
            &  values for the super learner ensemble.\\
    \bottomrule
  \end{tabular}
\end{table}

\section{Simulations}
\label{sim}

We developed a set of simulations where we varied the number of exposed subjects ($n\in \{200,500,1000\}$) and unexposed `controls' per case for each  $n$ ($C\in \{1,2\}$), which led to six  total sample sizes ($N \in\{400,600,1000,1500,2000,3000\}$) for our cohorts. We also drew random samples  of the same total sample size $N$ to provide a performance comparison to sampling that is not conditional on exposure. These random sample cohorts do not need observation weighting in the propensity score function for the parameter to be identifiable. 
The design of the simulated population 
had six baseline covariates $\mathbf{X}= (X_b:b=1,\ldots,6)$ with each $X_b \sim Bern(p_b)$ and $p_b=(0.6, 0.4, 0.4, 0.5, 0.4, 0.5)$. Exposure $E$ was dependent on  all six baseline covariates with $E \sim Bern(p)$ and $p={\text{logit}^{-1}}(3.0X_1+1.1X_2+2.2X_3-1.7X_4-4.8X_5-3.7X_6)$. The probability of exposure in this simulated population was 0.3712. Note that when $C=2$, the ratio of exposed and unexposed subjects in the cohort is similar, although not identical, to that of the source population.

We implemented propensity score function estimators using observation weights given by $w=0.37$  and $(1-w)\frac{1}{C}$ for the various sample sizes.   We also considered an unweighted stacked propensity score function and two weighted stacked propensity score functions that used  `$w$ plus error' to illustrate  settings where $w$ is estimated and not known with certainty.  A substantial amount of error was introduced for the `$w$ plus error' sensitivity analyses, and we imposed 10\% error in both directions for $w-10\%= 0.33$ and $w+10\%=0.41$. 
\begin{table}[t]
\small

  \label{sample-table2}
  \centering
  \begin{tabular}{lll}
    \toprule
    Learner     & R package     & Tuning \\
    \midrule
    Random forests  & \texttt{randomForest} \citep{liaw:2002}  & size=250     \\
    Classification tree      & \texttt{rpart} \citep{rpart} &       \\
    Logistic regression     &   \texttt{glm}  \citep{RCore}    &   \\
    Lasso penalized regression    &   \texttt{glmnet}  \citep{glmnet}      & \\
            Neural network     &   \texttt{nnet} \citep{MASS}     & size=2  \\    
                        Neural network     &   \texttt{nnet}      & size=3  \\   
                                    Neural network     &   \texttt{nnet}      & size=5  \\     
    \bottomrule
  \end{tabular}
    \caption{Learners in the Ensemble}
\end{table}

Seven learners, described in Table~\ref{sample-table2}, were ensembled to create stacked propensity score functions with the \texttt{SuperLearner} package \citep{Polley:vanderLaan13}. They were selected based on their ability to incorporate observation weighting -- not all algorithms and/or their \texttt{R} implementation have this functionality -- while still including a variety of learners that search the parameter space in different ways.  Libraries of algorithms larger than this demonstrative example are, of course, possible. The \texttt{SuperLearner} package itself allows for observation weights in the construction of the stacked algorithm. We performed 500 simulations for each sample size and estimator.

Evaluation of performance focused on bias, mean squared error (MSE), and relative efficiency. Bias was measured with respect to the true individual probability $p_i={P}^F(E_i=1\mid\mathbf{X}_i)$ assigned to each observation $i$ in the generation of $E$ described above.  Thus, we estimated bias as $\widehat{\text{bias}}=\lvert{P}^F(E=1\mid\mathbf{X})-\hat{P}^F(E=1\mid\mathbf{X})\rvert.$ Similarly, we estimated MSE with $\widehat{\text{MSE}}=[{P}^F(E=1\mid\mathbf{X})-\hat{P}^F(E=1\mid\mathbf{X})]^2.$
Relative efficiencies were calculated as $\widehat{\text{MSE}}_{\text{unweighted}}/\widehat{\text{MSE}}_{\text{weighted}}.$
These choices for evaluation metrics were driven by the need for \textit{consistent estimators} of the propensity score function for causal effect estimators that use this nuisance parameter.  Other evaluation metrics, such as balance on observables with percent standardized mean differences, are more suitable when the propensity score is used in the design phase to \textit{match observations to controls}, which is not the study design we consider here.  Area under the ROC curve is also a common metric  when \textit{accurate classification} is of primary concern, which is also not the priority here.  

\begin{figure}[p] 
   \centering
   \includegraphics[width=1\linewidth]{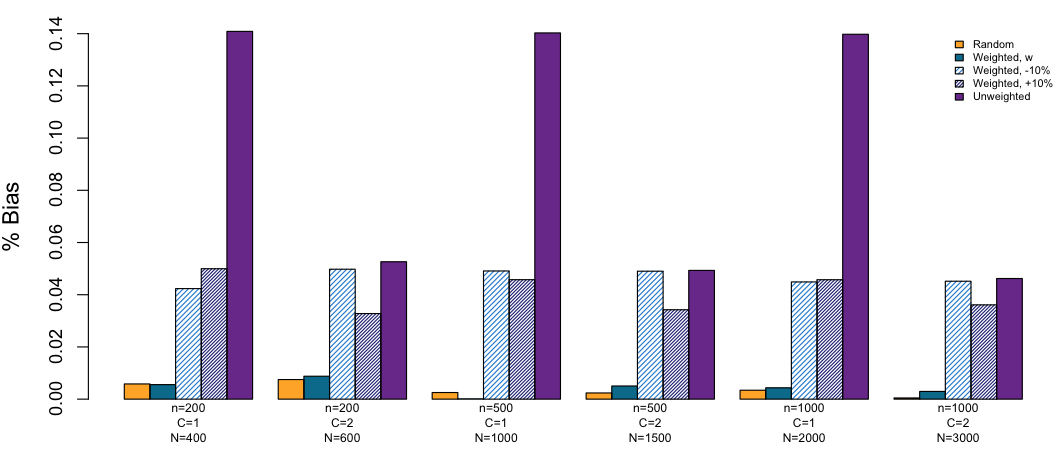}
   \caption{\small Percent bias plot for  ensembles in the random sample; three versions of the oversampled exposed subjects cohort \textit{with} observation weighting: $w$ and  `$w$ plus error' ($w-10\%$ and $w+10\%$); and oversampled exposed subjects cohort \textit{without} observation weighting. All values were computed across 500 simulations for each sample size and estimator.}
   \label{fig:sim1}
\end{figure}

\begin{figure}[p] 
   \centering
   \includegraphics[width=1\linewidth]{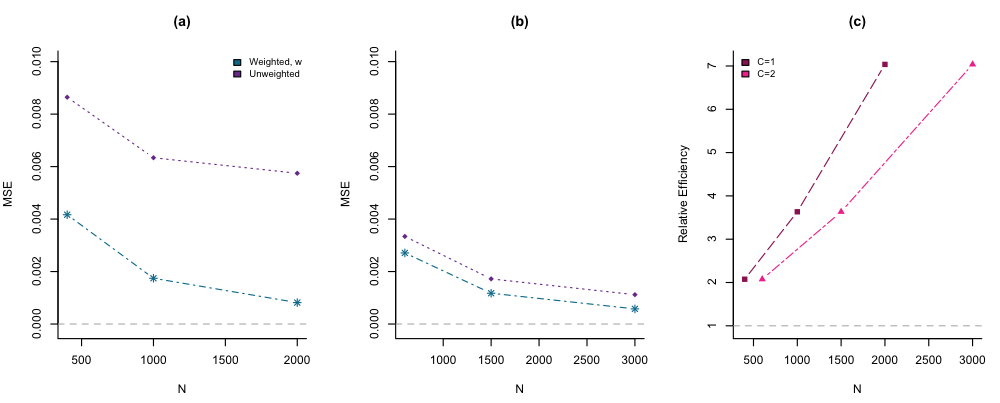}
   \caption{\small MSE plots for oversampled exposed subjects cohorts \textit{with} and \textit{without} observation weighting  across various sample sizes for (a) $C=1$ and (b) $C=2$. Results for the random sample are omitted as they were similar to and often overlapped with the oversampled exposed subjects cohort \textit{with} observation weighting. Similarly, results for observation weighting based on `$w$ plus error' are omitted due to overlap. Relative efficiency plot across sample sizes for both $C=1$ and $C=2$ displayed in (c). All values were computed across 500 simulations for each sample size and estimator.}
   \label{fig:sim2}
\end{figure}

Our results demonstrated that observation weighting with known $w$ had much lower bias than unweighted  propensity score functions for both $C=1$ and $C=2$ across various sample sizes, as seen in Figure~\ref{fig:sim1}. Bias for observation weighting with $w$ was always $<1\%$ while unweighted functions ranged from $5-14\%$. The bias of the $w$ weighted function was  similar to that seen in the randomly sampled cohort. The  propensity score functions with `$w$ plus error'  weights were still nontrivially less biased for ${P}^F(E=1\mid\mathbf{X})$ when $C=1$. When $C=2$, the functions with $w-10\%$ had similar bias to the unweighted function, although the functions with $w+10\%$ were all slightly less biased. 

MSE results in Figure~\ref{fig:sim2} show that while values decreased, as expected, as sample sizes increased, the unweighted  propensity score function MSE remained higher than the weighted functions in all settings. The difference in MSE levels was smaller for ${C=2}$, although relative efficiencies for the weighted propensity score functions increased as sample size increased for both $C=1$ and $C=2$. The  lasso penalized regression and logistic regression received the largest algorithm weights in all  stacked propensity score functions. This is not surprising given that the true functional form was a simple main terms  regression. Algorithms that searched the parameter space nonlinearly  received small algorithm weights ranging from 0.01 to 0.05.

\section{Health Policy Intervention}	

We briefly consider a simplified hypothetical policy intervention in an analysis of one of the largest health care claims databases. The  IBM MarketScan Research Databases are  a  collection of  insurance claims and enrollment information for privately enrolled individuals across the United States \citep{adamson2008health}.  The databases include millions of unique enrollees per year and contain demographic  and health condition variables, among others. Recent work in the policy literature has examined the impact of the Affordable Care Act \citep{pace2016early}, celebrity endorsements of medical testing \citep{desai2016celebrity}, and other non-randomized interventions using the  Marketscan database.

Our hypothetical health policy intervention is a new global payment system for coordinated care of patients, similar to the Alternative Quality Contract implemented in Massachusetts by Blue Cross Blue Shield or the  Medicare Accountable Care Organizations  launched by the Centers for Medicare and Medicaid Services. Suppose this global payment system is deployed for all privately insured enrollees in the New England census region, and privately insured enrollees from the other three census regions (Midwest, South, and West) were deemed to be a suitable control group. Evaluation endpoints under study might be health outcomes, quality, or spending, but we do not propose one here.

We sampled $n=5000$ subjects exposed to the hypothetical global payment intervention (i.e., recorded as living in New England, which received the payment reform, $E=1$) and $n\times C$ subjects who were not exposed to the intervention ($E=0$) for three settings: $C=1$, $C=2$, and $C=3$. The addition of $C=3$ was driven by an interest to explore a ratio of exposed and unexposed subjects in the cohort that was closer to that of the source population. All subjects were continuously enrolled for at least two years from 2011-2012, and 77 baseline variables $\textbf{X}$ included sex, two age categories (`age 21 to 34' and `age 35 to 54,' with `age 55 to 64' as the reference group), and 74 health  variables.   These health variables were the hierarchical condition categories (HCCs) frequently used for health care risk adjustment, and were constructed using International Classification of Diseases codes \citep{Popeetal11,kautter2014hhs}. The most prevalent HCCs in our cohorts with oversampled exposed subjects were major depressive and bipolar disorder (3\%); breast cancer (among age 50+) and prostate cancer (1\%); and heart arrhythmias (1\%). All other HCCs were less than 1\%. The probability of exposure was 0.1947 in the source MarketScan population of over 10 million enrollees who met our inclusion criteria. 

\begin{figure} 
   \centering
   \includegraphics[width=1\linewidth]{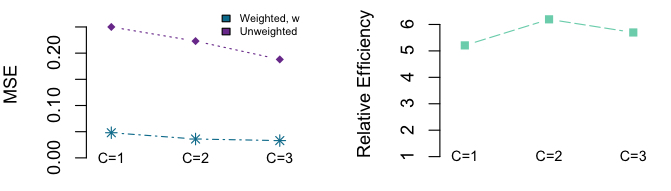}
   \caption{\small MSE and relative efficiency plots for $C=1$, $C=2$, and $C=3$ in a data analysis of oversampled exposed subjects cohorts \textit{with} and \textit{without} observation weighting.}
   \label{fig:data}
\end{figure}

We used the same library of learners as described in Section~\ref{sim}, and observation weights for exposed and unexposed were $w=0.19$ and $(1-w)\frac{1}{C}$. Given that we do not know the `true' probability of assignment to exposure as we did in our simulation experiments, we examine cross-validated MSE with respect to observed assignment to the exposed or unexposed group as well as relative efficiencies of these MSEs. Other metrics could be considered.  As in the simulations, we found our proposed observation weighted approach had smaller MSE and  was more efficient than the unweighted approach (at least five times as efficient for all $C$), shown in Figure~\ref{fig:data}. In all stacked propensity scores in this data analysis, the  classification tree algorithm (\texttt{rpart}) received the largest algorithm weight, ranging from 0.75 to 0.89.

\section{Concluding Remarks}

This paper proposed observation weights based on the probability of  exposure that can be used with a variety of algorithms in  propensity score functions for cohort studies with oversampled exposed subjects. We found this technique outperformed unweighted estimators and performed as well as randomly sampled cohorts, with respect to bias and variance. In sensitivity analyses using weights estimated with  error, performance was  substantially better than the unweighted  propensity scores when the proportion of exposed subjects in the sample was not close to that of the target population.  Interestingly, our results are suggestive that as the ratio of exposed and unexposed subjects approaches the true ratio seen in the source population, weighting may be less necessary, but potentially still provide benefits. The flexibility of this technique for cohorts with oversampled exposed subjects has strong potential for methods that use the propensity score function as a component in the estimator.  Thus, an immediate area of future work in cohort studies with oversampled exposed subjects is to develop and study the validity and efficiency of causal effect estimators, such as IPWs and double robust machine learning, when they incorporate weighted  propensity score functions. This includes accounting for the uncertainty of  estimated weights into the standard errors of the causal effect estimators.

\section*{Code}
Code for the simulated data and a demonstration of the computational implementation of the method is  online: \url{https://github.com/sherrirose/ConditionalCohortSamples}.

\bibliographystyle{Chicago}

\bibliography{TMLE_CI}
\end{document}